\documentclass[conference]{IEEEtran}

\usepackage{times}

\usepackage{amsmath}
\usepackage{amssymb}
\usepackage{graphicx}
\usepackage{textcomp}
\usepackage{subfigure}

\def\mvp{\vspace*{-0.1in}}

\IEEEoverridecommandlockouts

\begin{document}


\title{The {\em GangaWatch} Mobile App to Enable Usage of Water Data in Every Day Decisions Integrating Historical and Real-time Sensing Data}
\author{\IEEEauthorblockN{Sandeep S Sandha*\thanks{*This work was done while 
the author was at IBM Research - India.}}
\IEEEauthorblockA{UCLA\\
Los Angeles, USA}
\and
\IEEEauthorblockN{Biplav Srivastava}
\IEEEauthorblockA{IBM T J Watson Research Center\\
Yorktown Heights, NY, USA}
\and
\IEEEauthorblockN{Sukanya Randhawa}
\IEEEauthorblockA{IBM Research - India\\
Bangalore, India}}


\maketitle


\mvp
\mvp

\begin{abstract}
We demonstrate a novel mobile application called {\em GangaWatch}  that makes water pollution
data usable and accessible focusing on one of the most polluted river basins in the world. 
It is intended to engage common public who want to see water condition 
and safe limits, and their relevance based on different purposes. The data 
is a combination of old data determined from lab tests on physical samples and new data
from real time sensors collected from Ganga basin. The platform is open for contribution
from others, the data is  also available for reuse via public APIs, and it has 
already been used to derive new insights.
\end{abstract}


\section{Introduction}

Water is  important to a wide spectrum of everyday decision makers like farmers, tourists, environmentalists, health officials, policy makers and businesses due to its
unique roles as a life preserver. They and many more can benefit from decision support aids (i.e., AI systems) which can help them understand water pollution data and alternative decision choices that they may have. However, if one is looking for quality pollution data, one is lost. This is surprising given that there is a rich history in many countries of field-visits for data collection and lab-based testing, and they look forward to adopting real-time water sensing in a big way. A few  available 
are WaterLive mobile app for NSW, 
Australia \footnote{http://www.water.nsw.gov.au/realtime-data} and bath web app for 
UK\footnote{https://environment.data.gov.uk/bwq/profiles/} but they do not have  
rich features like multi-sensor data, support for multiple purposes and 
parameter range explanations.

We are particularly interested in the context of a developing country like India. The water of Ganga basin, her major river and its numerous tributaries, is used by over 400 million people, which is roughly one-third of India's population and equivalent to that of United States. Unfortunately, the state of Ganga is quite poor and data about its water pollution is not easily available.

GangaWatch app \cite{gw-app} fills the void by making water data usable and accessible to people. It is an experimental mobile application to show condition of water in the Ganga basin and beyond. It is intended for common public who want to see water condition and safe limits, as well as relevance based on different purposes. 
\section{ GangaWatch Details}
GangaWatch uses new real-time sensing data we collected from water bodies in 2015 and 2016. A unique aspect of the procedure followed was that three different techniques were used to assess water: lab samples were collected, real-time sensors were deployed and finally a mobile 
app\cite{bw-site} was used to record water condition, making correlation for data validation feasible. The app also gratefully uses open public data from sites like Indian open data which promote reuse, and more can be added by anyone using a public web or programmatic interface. In all, data from more than
60 locations in the Ganga basin is available in the app though not all is of the same volume
or has full range of values.
Information about safety limits and purposes are taken from guidelines by concerned
regulators - Central Pollution Control Board (CPCB, the federal pollution regulator) and 
Bureau of Indian Standards (BIS, national standards body).

Further, when monitoring water, a number of parameters are of
interest to different stakeholders depending on their purpose.
Environment agencies like EPA in US and CPCB in
India recommend tens of parameters - CPCB recommends
more than 30 parameters (\cite{cpcb-manual},\cite{cpcb-mon}) in
India.  Agencies
also recommend standards for different usage like drinking,
irrigation - CPCB prescribes parameters of interest and
their ranges for 22 industries (\cite{cpcb-mon}). A major challenge
is overlapping specifications of multiple agencies
within a government and also at multiple levels (national,
state and international) which can be in conflict. We reconcile
ranges for 25 parameters in the data released using
CPCB guidelines on purpose and pollutants.

\begin{figure}

\begin{tabular}{cc}
\includegraphics[angle=0,width=0.2\textwidth]{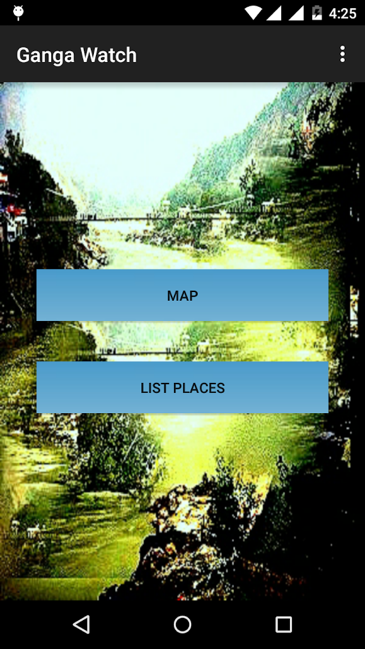}
&
\includegraphics[angle=0,width=0.2\textwidth]{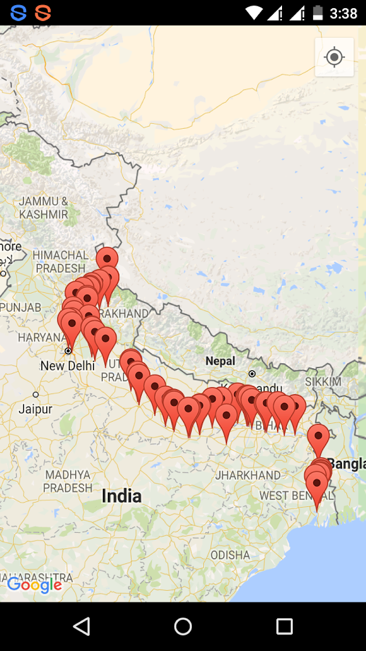}

\\

\includegraphics[angle=0,width=0.2\textwidth]{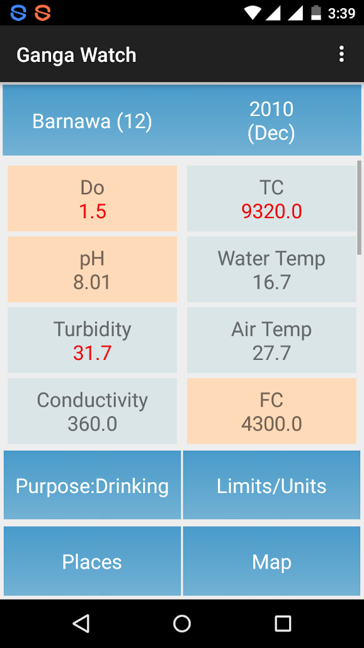}
&
\includegraphics[angle=0,width=0.2\textwidth]{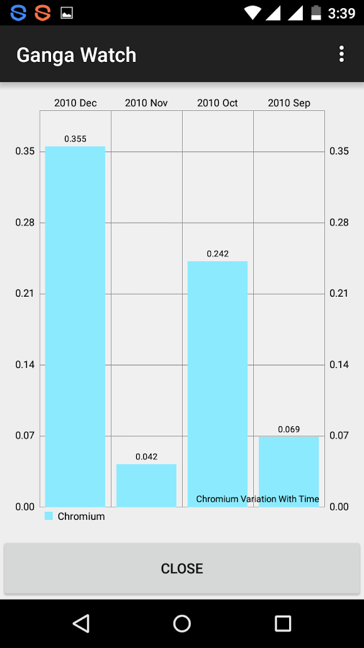}
\\

\includegraphics[angle=0,width=0.2\textwidth]{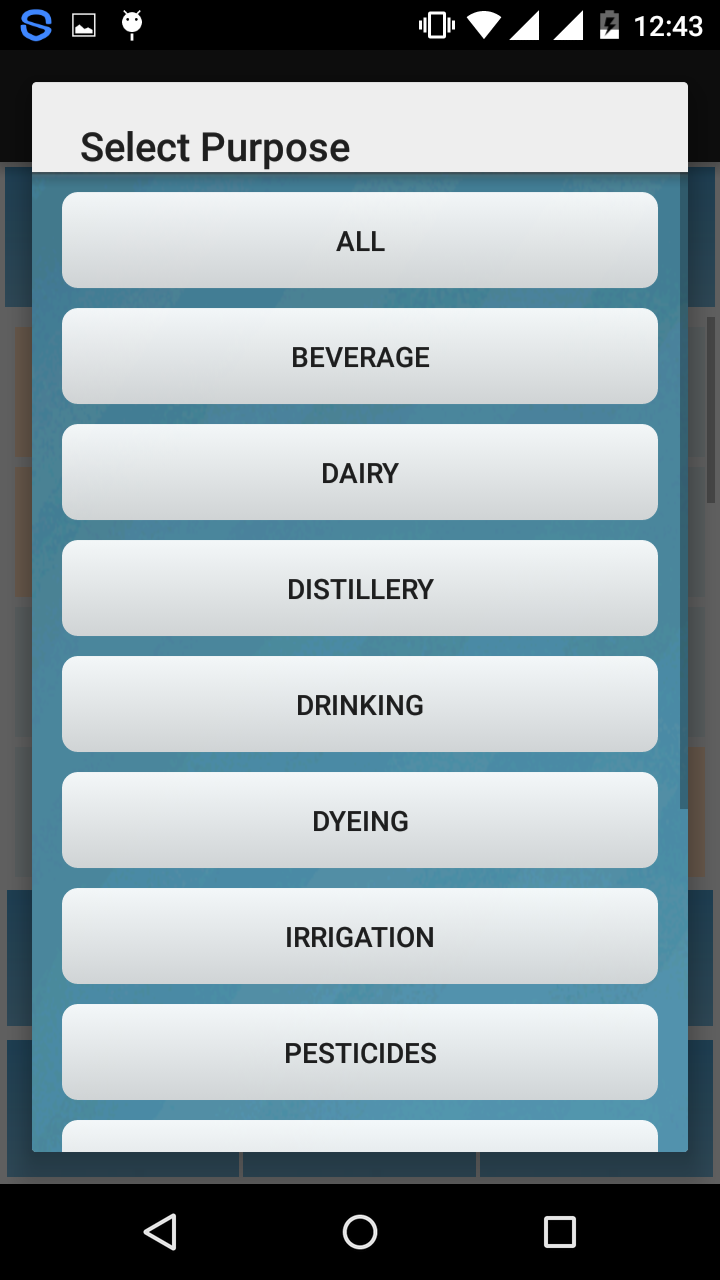}
&
\includegraphics[angle=0,width=0.2\textwidth]{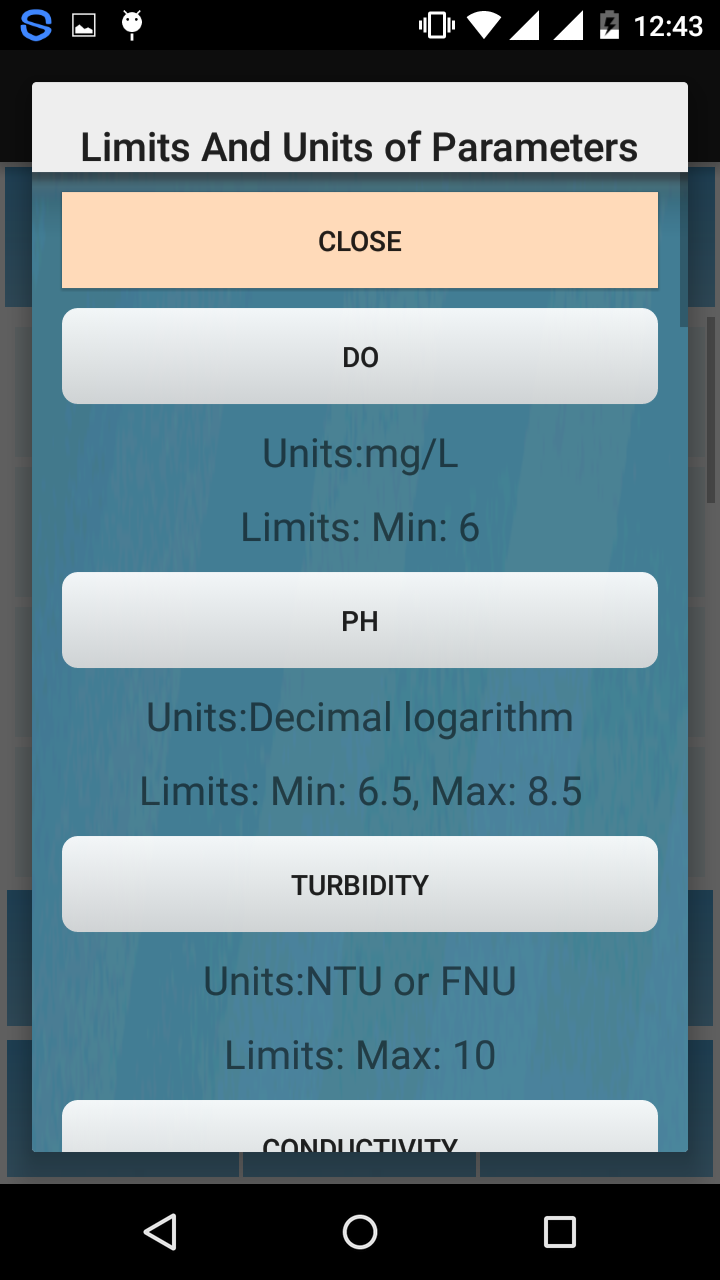}

\end{tabular}

\caption{ Screen-shots of GangaWatch. Views show opening the app, selecting a location, 
viewing relevant parameters and unsafe levels relevant to a purpose, 
looking at variations of a parameter over time, selecting alternative purposes
and understanding safety levels, respectively.} 
\label{fig:screen} 
\end{figure}

Figure~\ref{fig:screen}   shows screen shots of the app. When it is launched, the user
can select  a location of interest either on a map or from a list. The map
view shows all the locations where data is available but focusing particularly on the Ganga basin.
When the user selects a location, she can view all parameters supported in the tool
with their reported values.
Furthermore, parameters which are relevant to the (selected or default) purpose are
highlighted and those exceeding (or out-of-range from)  acceptable limits are colored
in red. In the example, {\em Drinking} purpose has parameters DO-Dissolved Oxygen,
pH, FC-Faecel Coliform and Chromium (not visible) highlighted. The user can 
select a parameter to visualize its changes over time, explore 
alternative purposes and also understand safe ranges for all parameters.

Serving the GangaWatch app is a cloud-based infrastructure called BlueWater \cite{bw-paper}. 
Here, data is stored in a commercial NO-SQL database (specifically Cloudant) 
on a server while it is accessed and manipulated using public APIs, and through them, 
possibly by any mobile and web client including GangaWatch.
The server-side stores, processes, and manages the collected data. 
 A secured web based portal is used for data upload. The demonstration will show the features of GangaWatch app, upload of a sample dataset 
as well as the working of the public APIs \cite{bw-site}.

A common issue which arises in storing sensor data is that of selecting the right data model.
Since we wanted to accommodate a wide diversity in water data in terms of temporal and 
spatial sampling rate, collection processes, pollution parameters and end-use applications, 
we decided  to record all available metadata, only standardizing on time and space (i.e., latitude, longitude, altitude) representations. This allows the platform to be expressive but requires
the application to query and choose the right metadata for its suitable analysis. We believe
this is the right middle-ground for a platform that aims to serves multiple applications.

\section{Discussion}
Water data from BlueWater can be used to demonstrate case studies of
data-driven decision making.  One example is to understand the impact of river-based tourism
when large number of people gather during religious festivals to bathe in the river (preliminary 
results under review).
Another example is to generate randomized inspection plans to monitor polluting 
industries on river banks \cite{nectar}. We hope the research community can use available data
to create more scenarios as well as contribute more data, in future.

Further, the app is not specific to any river basin. The only variabilities regions have
are in the list of purposes, parameters and safety ranges  one region of the world may 
prefer over the other. This can be supported in the current app itself, or a new one be created
but still using water data from BlueWater using public APIs.

%
%
%



\bibliographystyle{IEEEtran}
\bibliography{gw-demo}

\begin{thebibliography}{1}
\providecommand{\url}[1]{#1}
\csname url@samestyle\endcsname
\providecommand{\newblock}{\relax}
\providecommand{\bibinfo}[2]{#2}
\providecommand{\BIBentrySTDinterwordspacing}{\spaceskip=0pt\relax}
\providecommand{\BIBentryALTinterwordstretchfactor}{4}
\providecommand{\BIBentryALTinterwordspacing}{\spaceskip=\fontdimen2\font plus
\BIBentryALTinterwordstretchfactor\fontdimen3\font minus
  \fontdimen4\font\relax}
\providecommand{\BIBforeignlanguage}[2]{{%
\expandafter\ifx\csname l@#1\endcsname\relax
\typeout{** WARNING: IEEEtran.bst: No hyphenation pattern has been}%
\typeout{** loaded for the language `#1'. Using the pattern for}%
\typeout{** the default language instead.}%
\else
\language=\csname l@#1\endcsname
\fi
#2}}
\providecommand{\BIBdecl}{\relax}
\BIBdecl

\bibitem{gw-app}
I.~B. Machines, ``Gangawatch app on android store,'' in
  \emph{https://play.google.com/store/apps/details?id=com.ibm.
  research.gangawatch, Video: https://youtu.be/ MbVvVGsZoTo}, 2016.

\bibitem{bw-site}
------, ``Blue water site,'' in
  \emph{http://researcher.watson.ibm.com/researcher/ view\_group.php ?id=6924},
  2016.

\bibitem{cpcb-manual}
C.~P.~C. Board, ``Guide manual: water and wastewater analysis,'' in
  \emph{http://cpcb.nic.in/upload/NewItems/NewItem\_17
  \_guidemanualw\&wwanalysis.pdf}, 2014.

\bibitem{cpcb-mon}
------, ``Guidelines for online continuous monitoring system for effuents,'' in
  \emph{http://www.inpaper.com/FinalGuidelinse.pdf}, 2014.

\bibitem{bw-paper}
S.~S. Sandha, S.~Randhawa, and B.~Srivastava, ``Blue water: A common platform
  to put water quality data in india to productive use by integrating
  historical and real-time sensing data,'' in \emph{IBM Research Report
  RI15002.}, 2015.

\bibitem{nectar}
B.~Ford, M.~Brown, A.~Yadav, A.~Singh, A.~Sinha, B.~Srivastava, C.~Kiekintveld,
  and M.~Tambe, ``Protecting the nectar of the ganga river through
  game-theoretic factory inspections,'' in \emph{Proc. of 14th Practical Appl.
  of Scalable Multi-agent Sys. (PAAMS), Sevilla, Spain, June 1-3}, 2016.

\end{thebibliography}

\end{document}